\documentclass[12pt,preprint]{aastex}
\usepackage{graphicx}
\usepackage{amsmath}
\usepackage{natbib}
\usepackage{longtable}

\shorttitle{Polarization from BHs}
\shortauthors{Schnittman {\it et al.}}

\begin{document}

\title{X-ray Polarization from Black Holes: \\
{\it GEMS} Scientific White Paper}
\author{Jeremy Schnittman\altaffilmark{1},
Lorella Angelini\altaffilmark{1},
Matthew Baring\altaffilmark{2},
Wayne Baumgartner\altaffilmark{1},
Kevin Black\altaffilmark{1},
Jessie Dotson\altaffilmark{3},
Pranab Ghosh\altaffilmark{4},
Alice Harding\altaffilmark{1},
Joanne Hill\altaffilmark{1},
Keith Jahoda\altaffilmark{1},
Phillip Kaaret\altaffilmark{5},
Tim Kallman\altaffilmark{1},
Henric Krawczynski\altaffilmark{6},
Julian Krolik\altaffilmark{7},
Dong Lai\altaffilmark{8},
Craig Markwardt\altaffilmark{1},
Herman Marshall\altaffilmark{9},
Jeffrey Martoff\altaffilmark{10},
Robin Morris\altaffilmark{3},
Takashi Okajima\altaffilmark{1},
Robert Petre\altaffilmark{1},
Juri Poutanen\altaffilmark{11},
Stephen Reynolds\altaffilmark{12},
Jeffrey Scargle\altaffilmark{3},
Peter Serlemitsos\altaffilmark{1},
Yang Soong\altaffilmark{1},
Tod Strohmayer\altaffilmark{1},
Jean Swank\altaffilmark{1},
Yuzuru Tawara\altaffilmark{13},
and Toru Tamagawa\altaffilmark{14}}

\altaffiltext{1}{NASA/GSFC}
\altaffiltext{2}{Department of  Physics and Astronomy, Rice Univ.}
\altaffiltext{3}{NASA/ARC}
\altaffiltext{4}{Department of Astronomy and Astrophysics, TIFR, Mumbai, India}
\altaffiltext{5}{Department of Physics and Astronomy, University of Iowa}
\altaffiltext{6}{Department of Physics, Washington U. }
\altaffiltext{7}{Department of Physics and Astronomy, Johns Hopkins University}
\altaffiltext{8}{Department of Astronomy, Cornell University}
\altaffiltext{9}{Center for Space Research, MIT}
\altaffiltext{10}{Department of Physics, Temple U.}
\altaffiltext{11}{U. Oulu, Finland}
\altaffiltext{12}{Department of Physics and Astronomy, North Carolina State University}
\altaffiltext{13}{Nagoya University, Japan}
\altaffiltext{14}{Riken University, Japan}

\begin{abstract}
We present here a summary of the scientific goals behind the Gravity
and Extreme Magnetism SMEX ({\it GEMS}) X-ray polarimetry mission's
black hole (BH) observing program. The primary targets can be divided
into two classes: stellar-mass galactic BHs in accreting binaries, and
super-massive BHs in the centers of active galactic nuclei (AGN). The
stellar-mass BHs can in turn be divided into various X-ray spectral
states: thermal-dominant (disk), hard (radio jet), and steep power-law
(hot corona). These
different spectral states are thought to be generated by different
accretion geometries and emission mechanisms. X-ray polarization is an
ideal tool for probing the geometry around these BHs and revealing the
specific properties of the accreting gas.
\end{abstract}

\maketitle

\section{INTRODUCTION}
\label{intro}

The first positive detection of polarized X-rays from an astronomical
source was made with a sounding rocket experiment in 1971
\citep{novick:72}. The last positive detection of polarized X-rays
from an astronomical source was made with the Orbiting Solar
Observatory (OSO-8) in 1976 \citep{weisskopf:76}\footnote{More
  recently, there {\it has} been evidence of polarized hard X-rays and
  $\gamma$-rays from {\it INTEGRAL} and {\it Fermi}, but not in the
  2-10 keV band that {\it GEMS} will probe.}. Both observations
were of the Crab nebula, one of the brightest X-ray sources in the
sky, and quite highly polarized at a level of $\sim 20\%$. As
most polarization detectors are fundamentally limited by counting
statistics, it would take orders of magnitude more photons to reach
a sensitivity of a few percent, the level of polarization expected
from many classes of astrophysical sources. It is quite likely that,
in the era of OSO-8, no other source on the sky would have given even
a marginal detection. 

Now, after more than
thirty years since that last successful observation, we are on the
brink of a new era of discovery with X-ray polarization. A recent
flurry of new mission proposals has renewed interest in theoretical
modeling of X-ray polarization from a variety of astrophysical sources. The
Gravity and Extreme Magnetism SMEX ({\it GEMS}) mission\footnote{\tt
  heasarc.gsfc.nasa.gov/docs/gems}, with an expected launch date of
April 2014, will provide broad-band spectropolarimetry with high
sensitivity in the 2--10 keV band \citep{black:03, bellazzini:06, swank:09}.
{\it GEMS} should be able to detect polarization from a large
number of galactic and extra-galactic sources
at the $\delta \sim 1\%$ level, including stellar-mass black holes
(BHs), magnetars,
pulsar wind nebulae, and active galactic nuclei (AGN). Altogether,
there are a few dozen known X-ray sources that
are expected to have polarization and flux levels great enough to
be reliably detected by {\it GEMS}. This document outlines the primary
science questions regarding X-ray
polarization in accreting BHs, both the stellar-mass BHs
that are found in galactic X-ray binaries, and also the supermassive
BHs that power quasars and AGN at the centers of distant galaxies. 

A black hole is the quintessential compact object. Even the supermassive
BH in the center of our galaxy only subtends an angle of $\sim 2\times
10^{-5}$ arcsec. While radio interferometry is just now beginning to
achieve this level of angular resolution, X-ray telescopes are orders
of magnitude away, and unlikely to reach it in the foreseeable
future. Thus, to observe the detailed geometry of the X-ray emitting
regions of the accretion flow around a BH, we have to rely on indirect
measurements such as spectroscopy and timing, interpreted through
theoretical models for the source. Polarization will provide a new,
powerful tool for probing the geometry of BH accretion. At the most
basic level, polarization measures symmetry. A source with rotational
symmetry around the line of sight (e.g., an accretion disk viewed
face-on) will necessarily give zero net polarization. A source with
reflection symmetry across a plane (e.g., a non-relativistic disk
viewed at some angle) will be polarized either parallel or
perpendicular to that symmetry plane. In general, highly symmetric
objects will produce more highly-polarized signals than disordered or
turbulent systems. 

For extremely relativistic systems like BHs, the geometry of the
accretion flow is intimately mixed up with the geometry of the curved
space-time surrounding the BH. The effects of relativistic beaming,
gravitational lensing, and gravito-magnetic frame-dragging can break the
symmetry of even an ideal steady-state disk, and give
a non-trivial net rotation to the integrated polarization
vector.  Because the temperature in an accretion disk should
increase closer to the BH, where these relativistic effects are
strongest, it was predicted long ago that the observed angle and degree of
polarization of thermal disk emission should depend
on photon energy \citep{stark:77,connors:77,connors:80}. The
emissivity in the disk is itself a function of the space-time
geometry, since the spin of the BH strongly affects the efficiency of
angular momentum transfer and thus accretion, particularly in the
inner disk where temperatures are highest. 

To complicate
the matter further, many X-ray binaries and AGN exhibit strong
non-thermal radiation, strongly suggesting the presence of a hot
corona. The physical mechanism that produces this corona is not
currently well-understood, but is almost certainly magnetic in
origin. Additionally, many BHs produce strong relativistic jets of
similarly unknown origin. We hope that X-ray polarization will help us
constrain the many varyious models to explain these exotic systems. To
a large extent, the questions we want to answer are the same as those
that drive more established BH observations like spectroscopy and
timing analysis, but from the more geometric perspective of
polarization.

\section{BLACK HOLE STATES}
\label{BH_states}
Galactic BHs are known to exhibit remarkable variability in their flux
and spectral properties, on time scales from milliseconds to
years. \citet{remillard:06} identify four distinct states, based on
well-defined observational characteristics: Quiescent, Thermal
Dominant, Steep Power Law, and Hard. Since polarization measurements
require large numbers of photons, there is not much hope that we will
be able to observe sources in quiescence. The other three states are
likely associated with distinct accretion geometries, as described
below.
\vspace{-0.3cm}
\begin{center}
{\bf 2.1 Thermal Disk State}
\end{center}
\vspace{-0.3cm}

The simplest state is the Thermal Dominant state. This is defined by a
strong thermal peak in the spectrum, peaking around $1-3$ keV, with
little or no hard flux above $\sim 10$ keV. There is very little
variability, and no significant quasi-periodic oscillations
(QPOs). The physical model usually employed is that of a geometrically
thin, optically thick, steady-state accretion disk, aligned with the
BH spin axis. In the analytic ``alpha-disk'' model
\citep{shakura:73,novikov:73}, the gas orbits the 
BH on circular geodesic orbits and then plunges abruptly into the BH
at the inner-most stable circular orbit (ISCO). Since the ISCO is a
strong function of BH spin, any observable that depends on its
location provides a potential method for measuring spin. 

In recent years, a number of different groups have studied the
detailed properties of this plunging region using 3-D, fully
relativistic magneto-hydrodynamic (MHD) simulations. The results have
not been entirely conclusive. \citet{noble:08,noble:10} find
significant dissipation from gas near the ISCO, while
\citet{reynolds:08} and \citet{shafee:08,penna:10} find a more abrupt cutoff, similar to
the idealized case of \citet{novikov:73}, with the peak flux coming
from somewhat outside the ISCO, and little
or no emission coming from the plunging region. In either case, the
inner edge for any single source appears to be fixed over a range of
luminosities \citep{gierlinski:04}. If we wish to measure
BH spin with X-ray observations, whether by using continuum fitting,
broad iron lines, or polarization, it is critical to understand the
detailed behavior of the gas near the ISCO. 

\begin{figure}[h]
\begin{center}
\scalebox{0.6}{\includegraphics*[52,450][410,720]{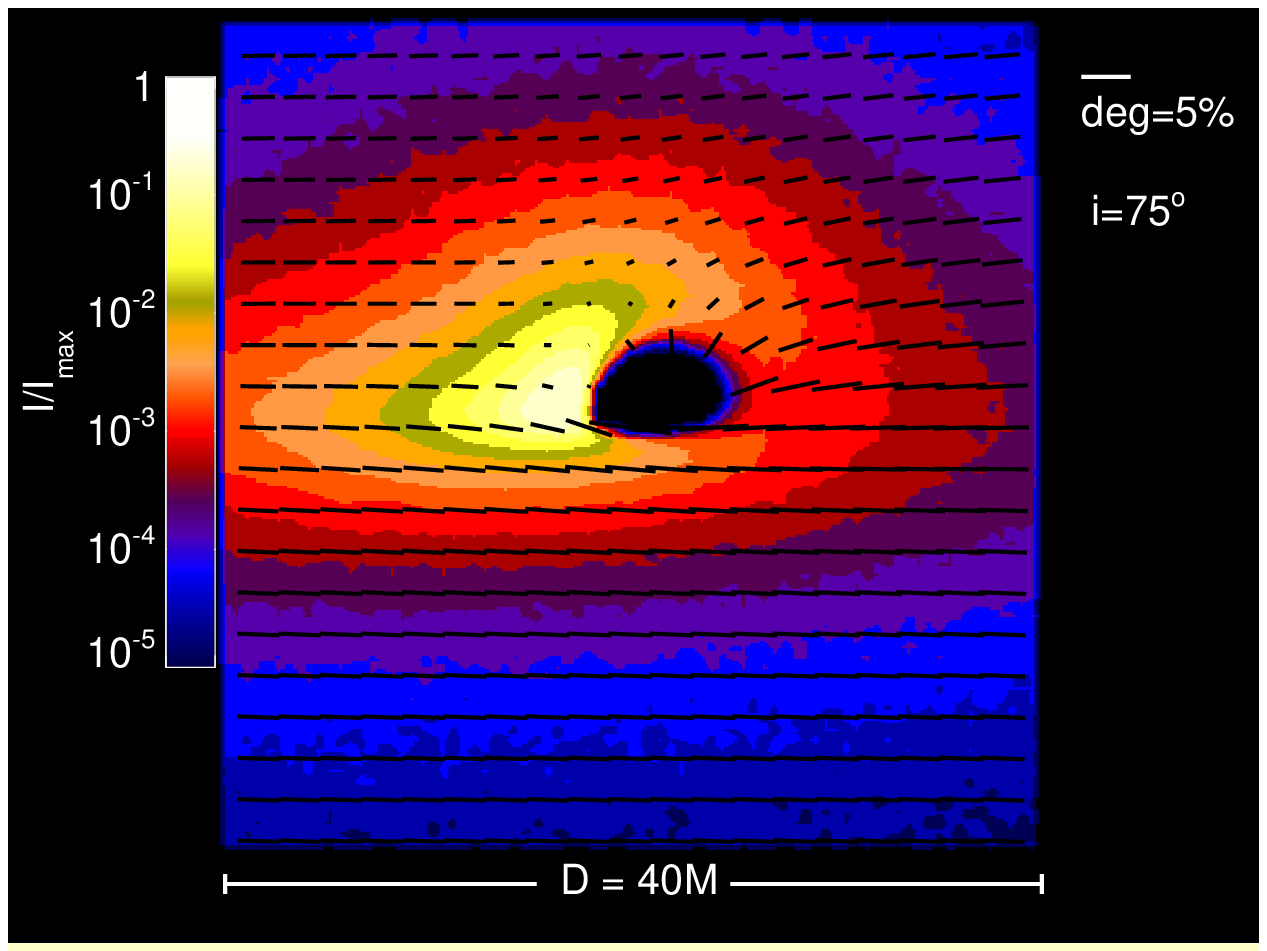}}
\scalebox{0.6}{\includegraphics*[52,450][410,720]{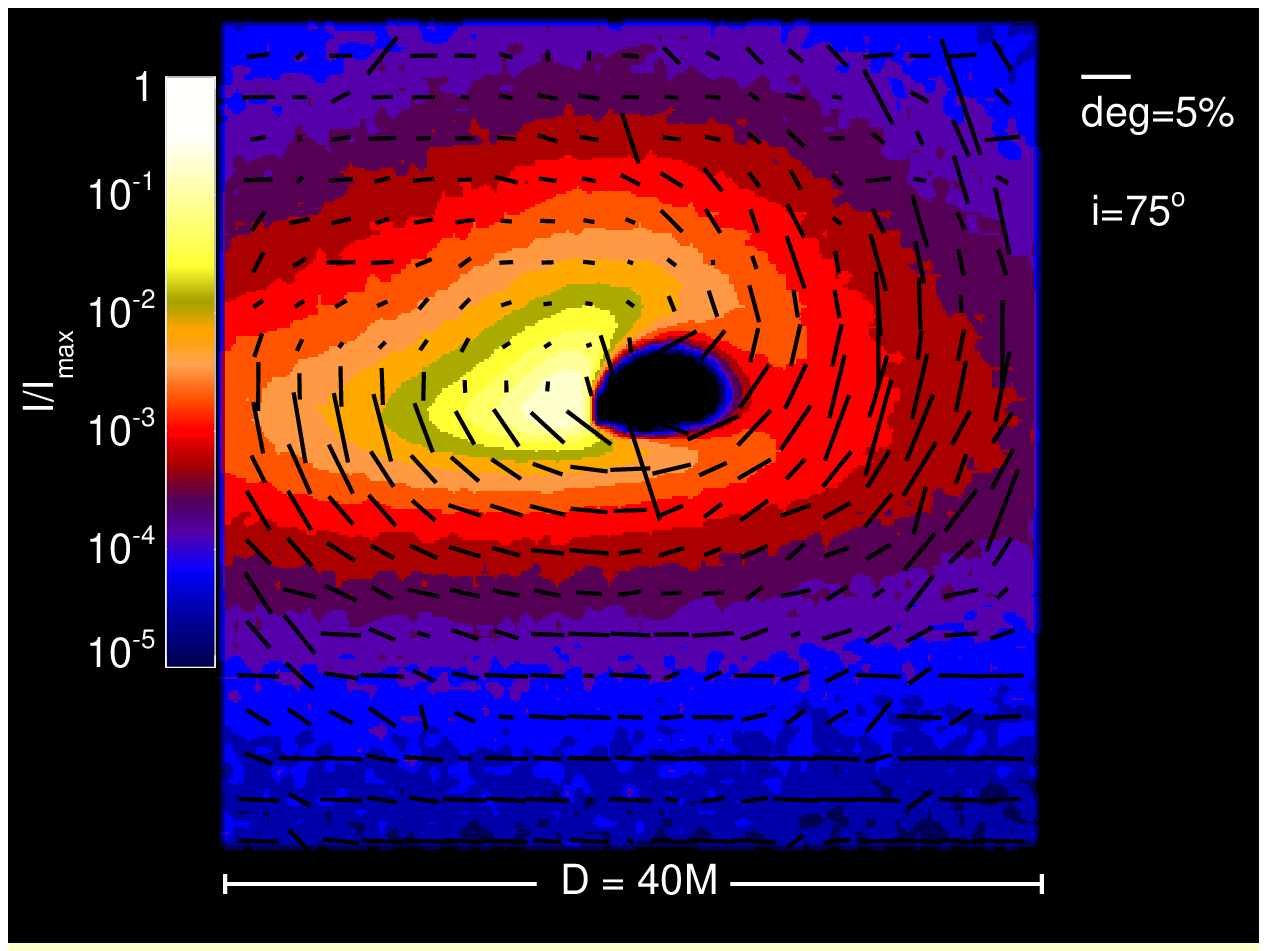}}
\caption{\label{thermal_image} \small Ray-traced image of direct radiation
  from a thermal disk ({\it left}). The observer is located at an inclination of
  $75^\circ$ relative to the BH and disk rotation axis, with the gas
  on the left side of the disk moving towards the observer, which
  causes the characteristic increase in intensity due to relativistic beaming.
  The black hole has
  spin $a/M=0.9$, mass $M=10 M_\odot$, and is accreting at $10\%$ of
  the Eddington limit with a Novikov-Thorne emissivity
  profile, giving peak temperatures around 1 keV. The observed intensity is color-coded on a
  logarithmic scale and the energy-integrated polarization vectors are
  projected onto the image plane with lengths proportional to the
  degree of polarization. The ({\it right}) image, includes the
  returning radiation, made up of photons emitted from the inner disk,
  deflected by the BH and scattered off the opposite side of the disk
  towards the distant observer. [reproduced from
  \citet{schnittman:09}] \normalsize}
\end{center}
\end{figure}

As shown in \citet{schnittman:09}, spectropolarization
observations of stellar-mass BHs in the thermal state may provide a
way to break the degeneracy between spin and magnetic torques inside
the ISCO. Polarization is particularly dependent on the emissivity and
orbital dynamics in the inner disk because of the strong relativistic
effects there. As originally pointed out in \citet{agol:00}, photons
emitted very close to the BH get strongly deflected by gravitational
lensing, and can actually pass over the BH and intersect with the far
side of the disk, a process known as ``returning radiation''
\citep{cunningham:76}. For highly ionized accretion 
disks (as expected for
the $\sim 1$ keV temperatures in typical X-ray binaries), the returning
radiation can scatter off the disk at large angles, which naturally
leads to high polarization. If even $5\%$ of the total flux returns to
the disk, it can dominate the polarization signal
\citep{schnittman:09}. Because the accretion disk opacity is expected
to be dominated by electron scattering, the seed polarization should
be emitted with modest polarization of a few percent, oriented
parallel to the disk surface in the local fluid frame \citep{chandra:60}. The scattered
radiation, on the other hand, will be highly polarized and
perpendicular to the disk surface, especially for observers at high
inclination angles. Figure \ref{thermal_image} shows polarization
images of a $10 M_\odot$ BH in the thermal disk state, including only
the direct radiation ({\it left}), and the total flux, including the
returning radiation ({\it right}).

\begin{figure}[h]
\scalebox{0.45}{\includegraphics{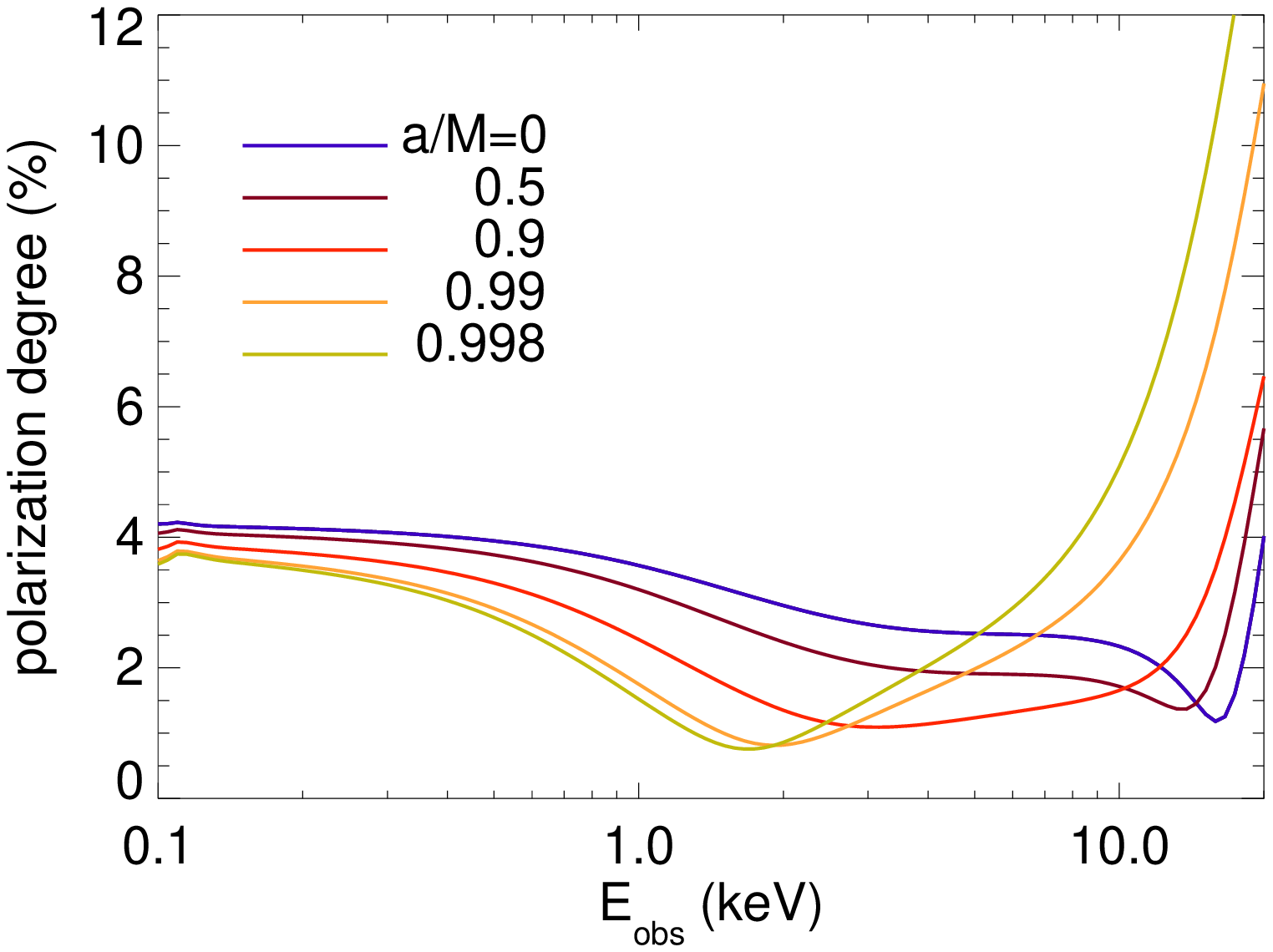}}
\scalebox{0.45}{\includegraphics{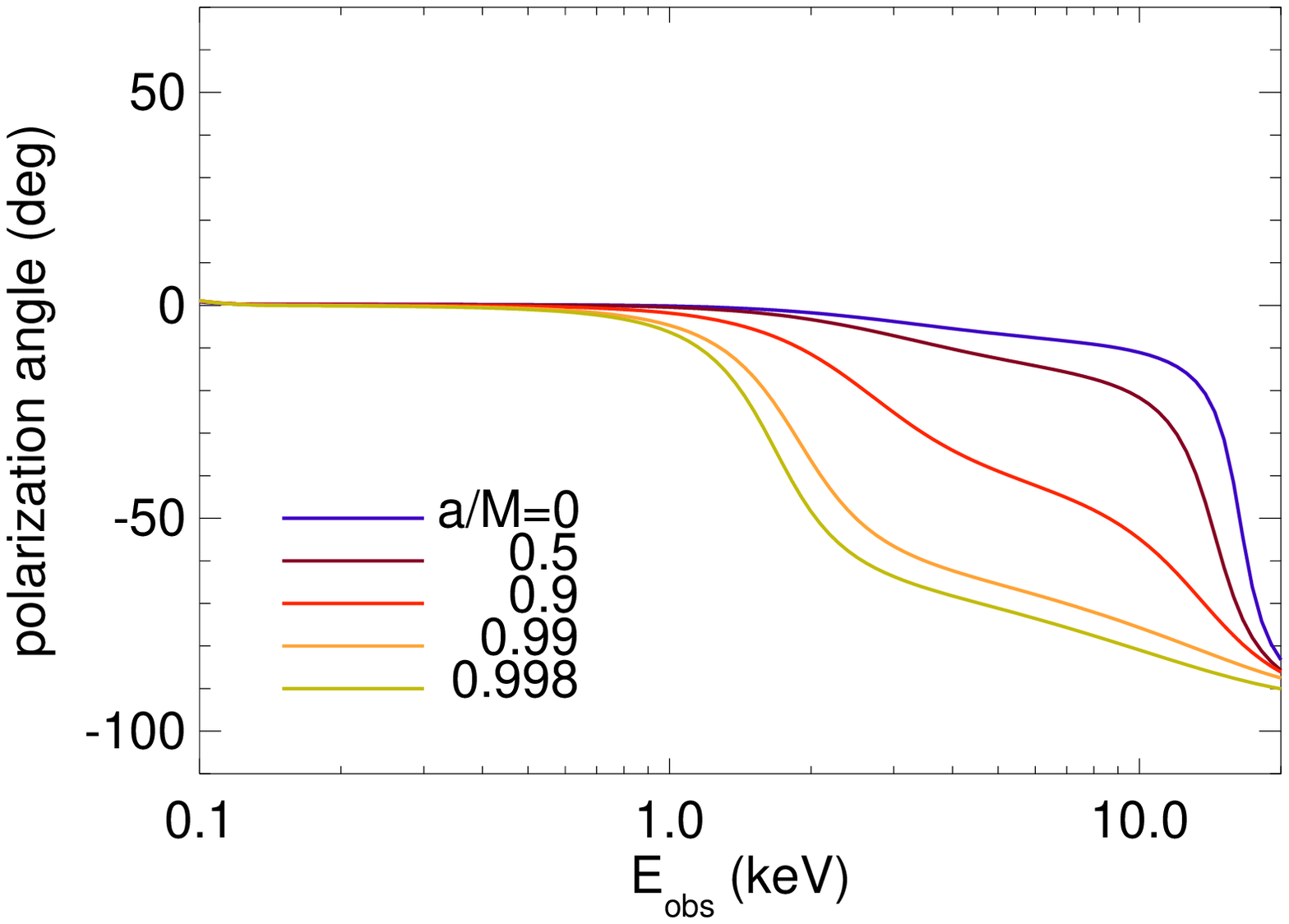}}
\caption{\label{thermal_spin} \small Polarization degree and angle
  for a range of BH spin parameters.
  All systems have inclination
  $i=75^\circ$, BH mass $10 M_\odot$, luminosity $L/L_{\rm Edd}=0.1$, and
  Novikov-Thorne radial emission profiles. [reproduced from
  \citet{schnittman:09} \normalsize} 
\begin{center}
\end{center}
\end{figure}

\vspace{-0.2cm}
Since the radiation emitted closest to the BH, where the disk is
hottest, is most likely to get deflected by the BH as returning
radiation, the high-energy flux tends to be polarized parallel to the
disk's rotation axis (``vertical'' in our convention, with $\psi=\pm
90^\circ$), while the low-energy flux from the outer disk is
oriented perpendicular to the disk axis, as projected on the sky
(``horizontal,'' with $\psi=0^\circ$). The
location and shape of this polarization transition is a direct probe
of the emissivity profile near the ISCO. Figure \ref{thermal_spin}
shows the polarization expected from a $10 M_\odot$ BH in the thermal
state, with a Novikov-Thorne emissivity profile that goes to zero at
the ISCO, for a range of spin parameters yet holding $L=0.1L_{\rm
  Edd}$ fixed. As the spin increases, more
of the flux originates closer to the horizon, and return radiation
becomes more important, leading to a larger fraction of observed
radiation in the vertical orientation. Aside from potentially
measuring the BH spin, this return radiation can be used to probe the
space-time as close to the horizon as possible and still allow the
photons to escape. In this context, the polarization swing from
horizontal to vertical might even be used to constrain alternative
theories of gravity, e.g., testing the no-hair theorem of general
relativity \citep{johannsen:10a,johannsen:10b,krawczynski:12b}.

\citet{schnittman:09} also
allowed for emission inside of the ISCO by including a simple
power-law parameterization that connected smoothly with the
Novikov-Thorne emissivity profile in the inner disk. Thus the
spectropolarization signal from any disk could be completely described
by only a few parameters: BH mass, spin, luminosity, distance,
inclination, and the additional power-law parameter for emission
inside the ISCO.

Even if such a simple model were a perfect description of the BH
source, there still exist fundamental degeneracies among the
parameters, in particular the mass, luminosity, and distance. These
degeneracies may be broken by observations at other wavelengths, e.g.\
optical absorption lines in the companion star that could be used
to make radial velocity measurements, in turn giving the binary mass
function. Other observational techniques, such as the continuum
fitting method, require {\it a priori} knowledge of the disk
inclination \citep{shafee:06}, while interpreting measurements of the broad iron
K$\alpha$ line typically is based on some assumption of the ionizing
flux distribution, and requires some knowledge of both the emission
and reflection edges of the accretion disk \citep{krolik:02}. X-ray polarization promises to be quite
complementary to these more established techniques. For example, the
polarization
below $\sim 1$ keV should probe the Chandrasekhar regime of the disk,
and thus provide a good estimate of the inclination, especially in
cases where the inner disk is tilted with respect to the orbit of the
binary system \citep{li:08}. Additionally, if the orientation of the
disk axis can be determined, we will be able to compare it with
observations of radio jets from the same souce \citep{reid:11}.
In \citet{schnittman:09} a simple data analysis method was used
to estimate our ability to recover the model parameters of
a hypothetical BH source with an idealized detector response
function. As expected, there was some degeneracy between the
BH spin and emissivity parameters, but with sufficient
signal-to-noise, these parameters may both be determined
independently. {\it Similar calculations must be repeated with the
  specific GEMS response function.}

Despite these theoretical uncertainties, the thermal disk state is
still arguably the most well-understood state for BH binaries, and
promises to give the most fundamental physical measurements, allowing
us to probe strong-field general relativity with a relatively small
uncertainty on the astrophysical factors that often plague such
investigations. The central questions we hope to address are:
\begin{itemize}
\vspace{-0.2cm}
\item What is the emissivity profile of optically thick, magnetized
  accretion disks near the BH, and to what extend will X-ray
  polarization be able to measure this profile?
\vspace{-0.2cm}
\begin{itemize}
  \item How does this profile depend on BH spin, and how accurately
    can we measure the spin by using polarization measurements?
\vspace{-0.2cm}
  \item Is the radiation generated in the inner disk and plunging
    region thermalized? 
\vspace{-0.2cm}
  \item Where is the inner edge of the disk, as defined by emission? 
\vspace{-0.2cm}
  \end{itemize}
\item What is the polarization of seed photons in a turbulent,
  magnetic (yet still thermal-dominated) accretion disk? 
\vspace{-0.2cm}
\item Can spectropolarization give definitive evidence for strong
  gravitational lensing? How well can we use the light bending of
  return radiation to probe strong-field gravity and test general
  relativity? 
\vspace{-0.2cm}
\end{itemize}

\vspace{-0.3cm}
\begin{center}
{\bf 2.2 Hot Corona State}
\end{center}
\vspace{-0.3cm}

The brightest BH transients usually fall into the category of the
``Very High'' or ``Steep Power Law'' (SPL) state, characterized by a
power-law spectrum with $I_\nu \sim \nu^{-\alpha}$, $(\alpha \gtrsim
1.4)$ in addition to a weaker thermal peak. The SPL 
state is also where we find most of the high-frequency quasi-periodic
oscillations (QPOs), which appear strongest in the non-thermal part of the
X-ray spectrum \citep{remillard:06}. One popular physical picture
behind the SPL state is one of a relatively cool thermal accretion
disk surrounded by a geometrically thick corona of hot electrons with
temperature $\gtrsim 50$ keV \citep{zdziarski:04,zdziarski:05}. From the slope of the power-law
spectrum, the Compton $y$-parameter for the corona is thought to be
of order unity, thus implying an optical depth of $\tau \sim 1-2$
\citep{rybicki:79}. Other suggested models include bulk Comptonization
from a converging accretion flow \citep{titarchuk:02,turolla:02}.

As evidenced by the plethora of models for the SPL, X-ray
spectroscopic observations have not been able to constrain the
detailed geometry of the corona very well in galactic BHs. Is it clumpy or
homogeneous? Is it centrally concentrated or diffuse? What is the
scale height? The answers to these questions should in turn provide
important clues about the physical origin of the corona. Is it
magnetically dominated like the solar corona? Is it gravitationally
bound to the BH? A better understanding of the coronal properties will
also lead to improved modeling of the broad iron emission line, which
is likely excited by the high-energy photons originating in the
corona. 

\begin{figure}[h]
\scalebox{0.45}{\includegraphics{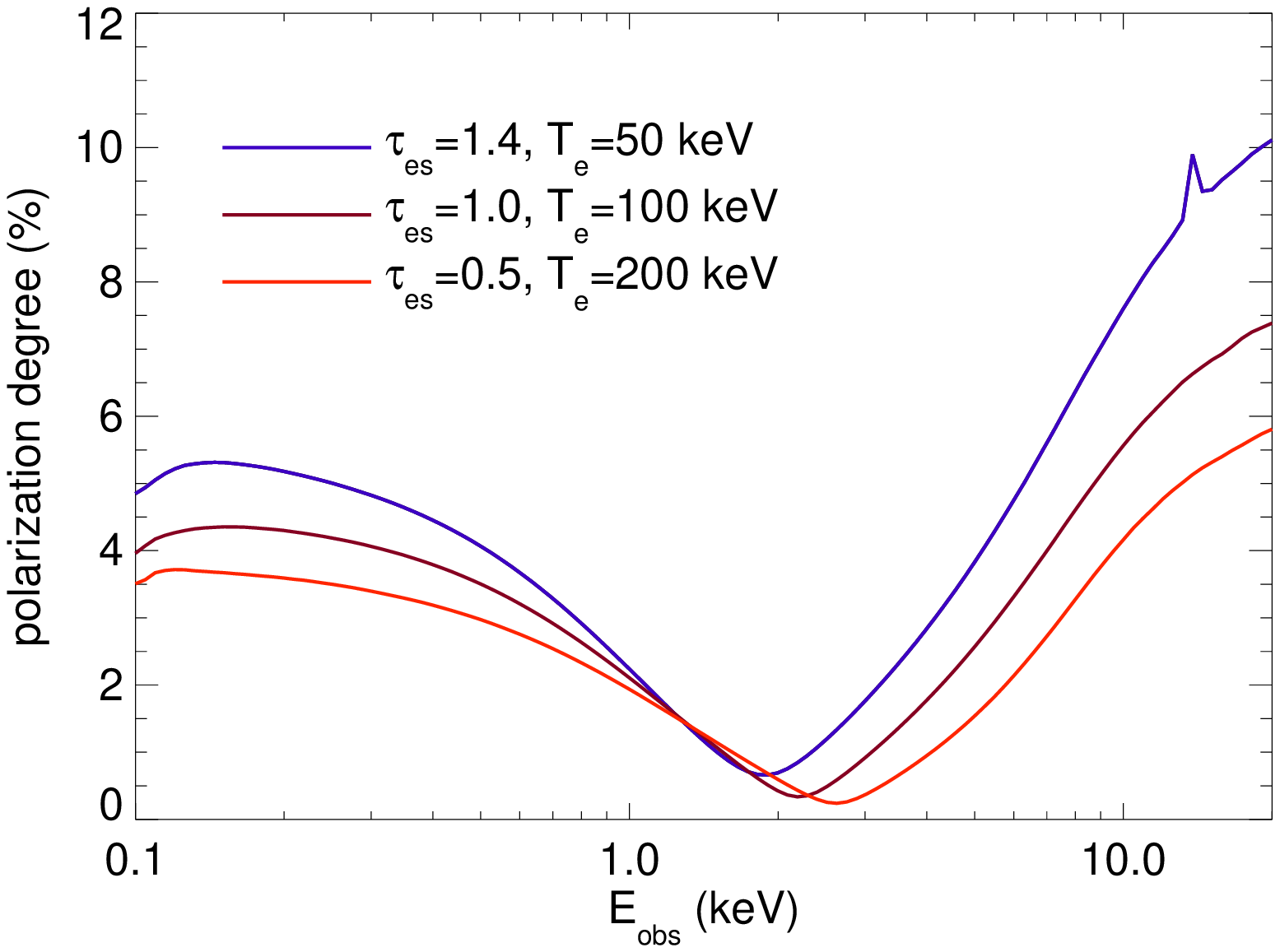}}
\scalebox{0.45}{\includegraphics{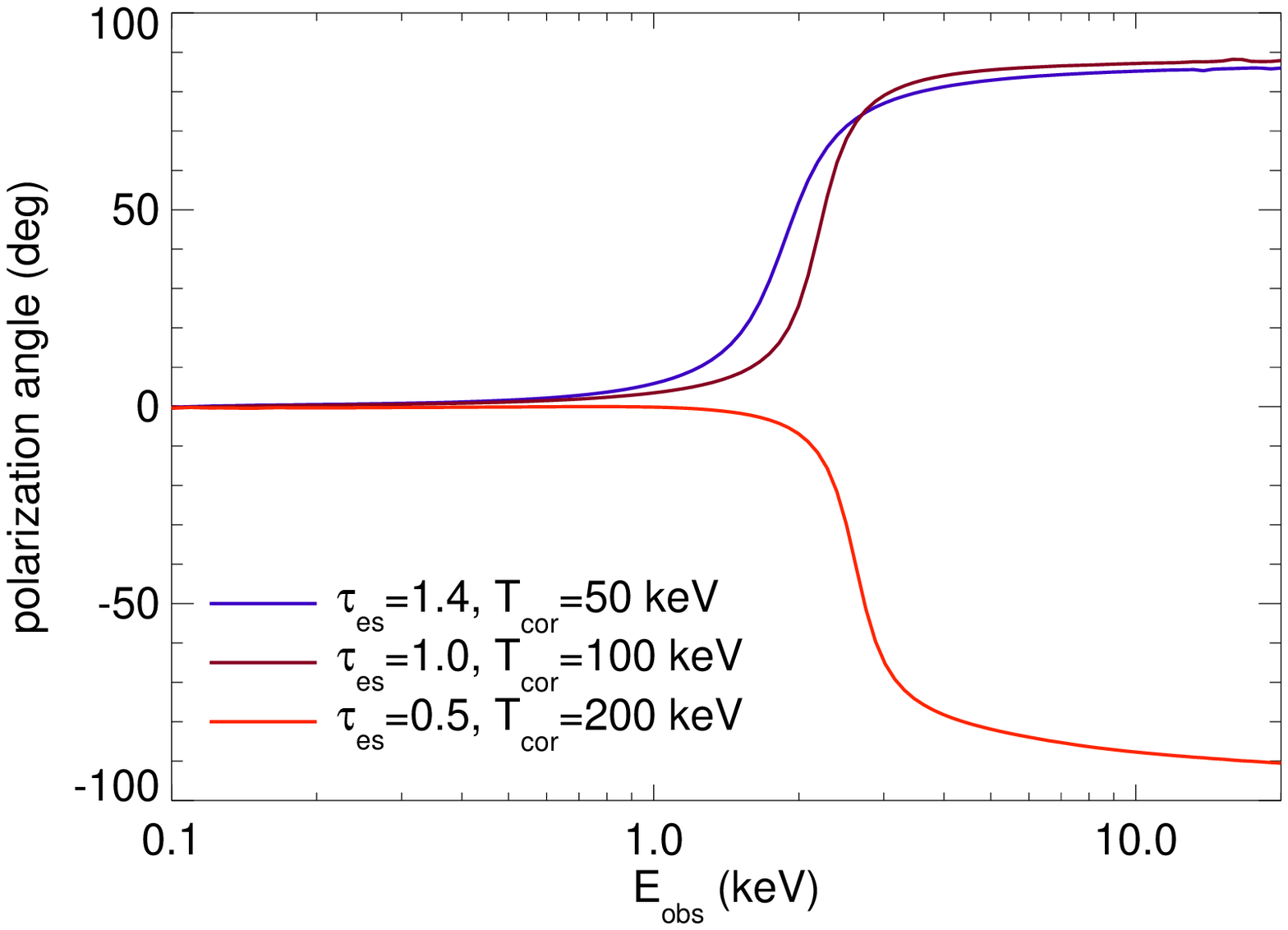}}
\caption{\label{wedge_tauT} Degree and angle of polarization for a
  sandwich corona, varying the optical depth and electron
  temperature, maintaining a roughly constant Compton-$y$
  parameter. [reproduced from \citet{schnittman:10}]}
\begin{center}
\end{center}
\end{figure}

\citet{schnittman:10} considered a few different simple
geometries for the corona and used a Monte Carlo ray-tracing code to
calculate the expected polarization signature for each case. Figure
\ref{wedge_tauT} shows one example of how polarization may provide
additional information about the properties of the corona. The
geometry considered here is a smooth sandwich with constant opening
angle of $H/R=0.1$, and vertically-integrated optical depth constant
across the entire disk. The same qualitative behavior is found as in
the thermal disk, with the polarization swinging from horizontal at
low energies to vertical at high energies. However, unlike the thermal
state where the transition is due to the global geometry of return
radiation, in the coronal state, the polarization transition can be
explained by more local effects. When the optical depth of the hot
corona is of order unity and the scale height is relatively small
($H/R \lesssim 0.1$), photons that scatter multiple times are
constrained to move in a plane parallel to the disk, and thus
are preferentially polarized in the vertical direction. The more
scattering events, the stronger the degree of polarization
\citep{sunyaev:85,matt:93,poutanen:96}. Since each
scattering event in the hot corona increases the photon energy by an
average fractional amount, those photons that scatter more will have
larger polarization and higher energy.

The sandwich
models shown in Figure \ref{wedge_tauT} were chosen to all have the
same Compton $y$-parameter and thus produce nearly identical
broad-band spectra. Yet the polarization signals clearly distinguish
between them. The reason is simple enough: to reach a certain energy,
the same seed photon would have to scatter more times in a warm corona
($T_{\rm c} \approx 50$ keV) than in a hot corona ($T_{\rm c} \approx
200$ keV). More scattering requires a more constrained geometry, which
in turn naturally leads to higher polarization.

In addition to the planar sandwich geometry, \citet{schnittman:10}
also considered
toy models such as a spherical central corona surrounded by a
truncated thermal disk, and a clumpy corona with a finite number of
hot blobs orbiting above the disk. In all cases, the characteristic
transition from horizontal to vertical polarization was present,
suggesting that low-resolution measurements made with the first
generation of X-ray polarimeters may not be able to distinguish
between these simple models. On the other hand, if we do {\it not} see such
a swing, or if the polarization is below $\sim 1-2\%$ across the
entire band, then
whole families of coronal models could be ruled out. 

In conjunction with more traditional X-ray observations like
spectroscopy and timing, polarization will help to address
these questions:
\begin{itemize}
\vspace{-0.2cm}
\item What are the physical properties of the corona: density,
  temperature, scale height, magnetization, homogeneity?
\vspace{-0.2cm}
\item What is the physical origin of the hot corona? 
\vspace{-0.2cm}
\item How does the hot corona interact with the cool disk; what is the
  nature of the boundary between these different regions?
\vspace{-0.2cm}
\item What is the profile of ionizing flux that generates iron
  flourescent lines; does it originate from the corona or disk or jet? 
\vspace{-0.2cm}
\item Where is the inner edge of the disk, as measured by reflection
  opacity? 
\vspace{-0.2cm}
\end{itemize}

For particularly bright galactic sources such as GRS 1915+105, even
first-generation instruments like {\it GEMS} should be able to make
some time-dependent observations of variable polarization. As one
example, we plan to model the time-varying polarization from the large
amplitude low-frequency ($\sim 1-10$ Hz) QPOs often seen seen in the
SPL state. One possible physical model for these QPOs is that of an
inclined disk or torus precessing around the spinning BH at the  
Lense-Thirring frequency, modulating the flux and
spectrum \citep{schnittman:06a,ingram:09}, and the polarization
as well. Since the inclination of the inner disk is changing with the
flux, polarization is an ideal tool for either confirming or
challenging this model. In GRS 1915+105, the $\sim$ 1 Hz QPOs are
strong enough to resolve and add in phase over many periods
\citet{miller:05}. 

\vspace{-0.3cm}
\begin{center}
{\bf 2.3 Hard State}
\end{center}
\vspace{-0.3cm}

\citet{remillard:06} define the Hard State as one where the spectrum
is dominated by a flat power-law with ($0.4 \le \alpha \le 1.1$) and no
significant contribution from a thermal disk. There are often strong
QPOs in the $1-10$ Hz frequency range, and also radio jets are quite
common \citep{fender:04}. As in the coronal SPL state discussed above,
there remain sizable uncertainties 
about the nature and geometry of the hard X-ray emitting region.
Popular models for the hard state of stellar-mass BHs include a
cool disk truncated at large radius ($\sim 100\, GM/c^2$)
surrounding a hot, radiatively inefficient flow
\citep{gierlinski:97,mcclintock:01,esin:01,done:09}, or
alternatively a more extended disk
surrounded by an optically thick hot corona, possibly in the form of a
hot wind \citep{blandford:04,miller:06,reis:09}.  
While the geometry of the corona could be as simple as
a uniform slab in the SPL state \citep{zdziarski:05}, the hard state
of galactic BH binaries,
as well as the X-ray emission from AGN, are more likely caused by
clumpy, inhomogeneous coronas, possibly caused by magnetic flares
\citep{haardt:94, poutanen:99}. Even in the cases where radio jets are
seen coincident with the X-ray observations (in either AGN or galactic
BHs), it is not clear if the
hard X-rays are coming predominantly from the jet
\citep{markoff:05,russell:10}, 
from the corona \citep{begelman:87,malzac:09}, or from some combination of the
two \citep{maitra:09}.

Models for coronae and truncated disks will be similar to those used
above for the SPL state. More theoretical work will need to be carried
out to develop comparable models for simulating the polarization from
jets. Some work in this area already exists in the AGN context
\citep{begelman:87, poutanen:94} and for one specific stellar-mass BH
picture \citep{mcnamara:09}; we need to generalize these models to a
wider range of galactic black holes in the hard state. 
Jets are known to produce highly polarized flux in
other wavelengths, generally attributed to synchrotron radiation of
hot electrons in a coherent magnetic field. These hot electrons can
produce synchroton self-Compton radiation in the X-ray
band, also strongly polarized \citep{poutanen:94}. The jet can also
serve as a source for hard X-rays that can get polarized by scattering
off the accretion disk/corona (the ``lamp post'' model;
\citet{dovciak:04}). Additionally,
low-energy seed photons from the disk may get upscattered in the jet,
further adding to the polarization signal (``external
Comptonization;'' \citet{mcnamara:09}). In either case, the total flux
from the jet may be relatively small, but highly polarized. With new models
for X-ray polarization from BH jets, we may be able to distinguish
between the coronal and jet models for the physical nature of the low
hard state. 

We have learned a great deal from polarization measurements of jets in
the radio, IR, and optical bands, but
because they can probe different length and energy scales, we expect
X-ray polarization measurements
to significantly improve our understanding of BH jets, helping to
answer the central questions: What is the magnetic field structure of
the jet? What are the basic physical properties of the jet: density,
composition (baryons vs pairs), Lorentz factor, electron temperature?
How is the jet generated near the BH? What role does BH spin
play in jet production and strength? What is the source of the seed
photons for inverse Compton processes? 

As we have seen with the coronal models, by adding more model parameters,
it often becomes more difficult to constrain any one of them with 
observations. Thus it will be critical to provide quantitative 
estimates of the confidence limits achievable by polarization
observations of real astrophysical sources. At the very least, we
expect that X-ray polarization detections (or even upper limits) will
be able to rule out entire classes of jet models. Here
multi-wavelength observations will be particularly valuable, using
radio observations to confirm the presence of a jet and possibly its
orientation \citep{fender:04}.

\section{ACTIVE GALACTIC NUCLEI}
\label{AGN}

It has long been known that a significant fraction of the flux
from AGN is emitted in the X-ray band
\citep{elvis:78}. As in stellar-mass BHs, this high-energy flux likely
comes from lower-energy seed photons inverse-Compton scattered in a
corona of hot electrons with $T_c \sim 100$ keV. Similar to the
stellar-mass case, this leads to a relatively hard power-law spectrum
with index $\alpha \sim 0.5-1$
\citep{nandra:91,mushotzky:93}. However, unlike the stellar-mass case,
the temperature of the inner disk for an AGN will be well below a keV,
leading to a thermal peak in the UV band. Furthermore, even when the
disk is dominated by radiation pressure and electron scattering
opacity, there should still be a substantial fraction of metals that
are not fully ionized, producing a large opacity for absorption
above $\sim$ 1 keV. 

Both of these AGN features---lower energy seed photons and an X-ray
absorbing disk---lead to important differences in the polarization
signature as compared to the stellar-mass case. Because the seed photons
start off with lower energies, they
must scatter more times in the corona in order
to reach the $\sim 1-10$ keV band. For a thin sandwich corona, this
means that the scattering geometry is even more constrained than in
the stellar-mass case, forcing the photons to move in a plane parallel
to the disk surface, leading to a stronger vertical
polarization \citet{poutanen:96}. The AGN disk absorbs much of the incident X-ray flux
from the corona, so the Compton $y$-parameter is effectively smaller than
that of a stellar-mass system with the same coronal properties because
scattering sequences are halted once a photon strikes the disk,
thereby reducing the average path length of the photons that escape to
infinity.  An absorbing disk boundary condition with a sandwich corona
therefore leads to an even higher degree of
X-ray polarization because the photons are forced to scatter in a more
constrained geometry before escaping the corona. 

\begin{figure}[h]
\caption{\label{agn} \small Degree and angle of polarization for a
  supermassive BH with $M=10^7 M_\odot$, $a/M=0.9$, and $i=45^\circ$,
  with corona temperature $T_c=100$ keV and varying covering
  fractions. The thermal disk flux peaks around 100 eV. \normalsize}
\begin{center}
\scalebox{0.45}{\includegraphics{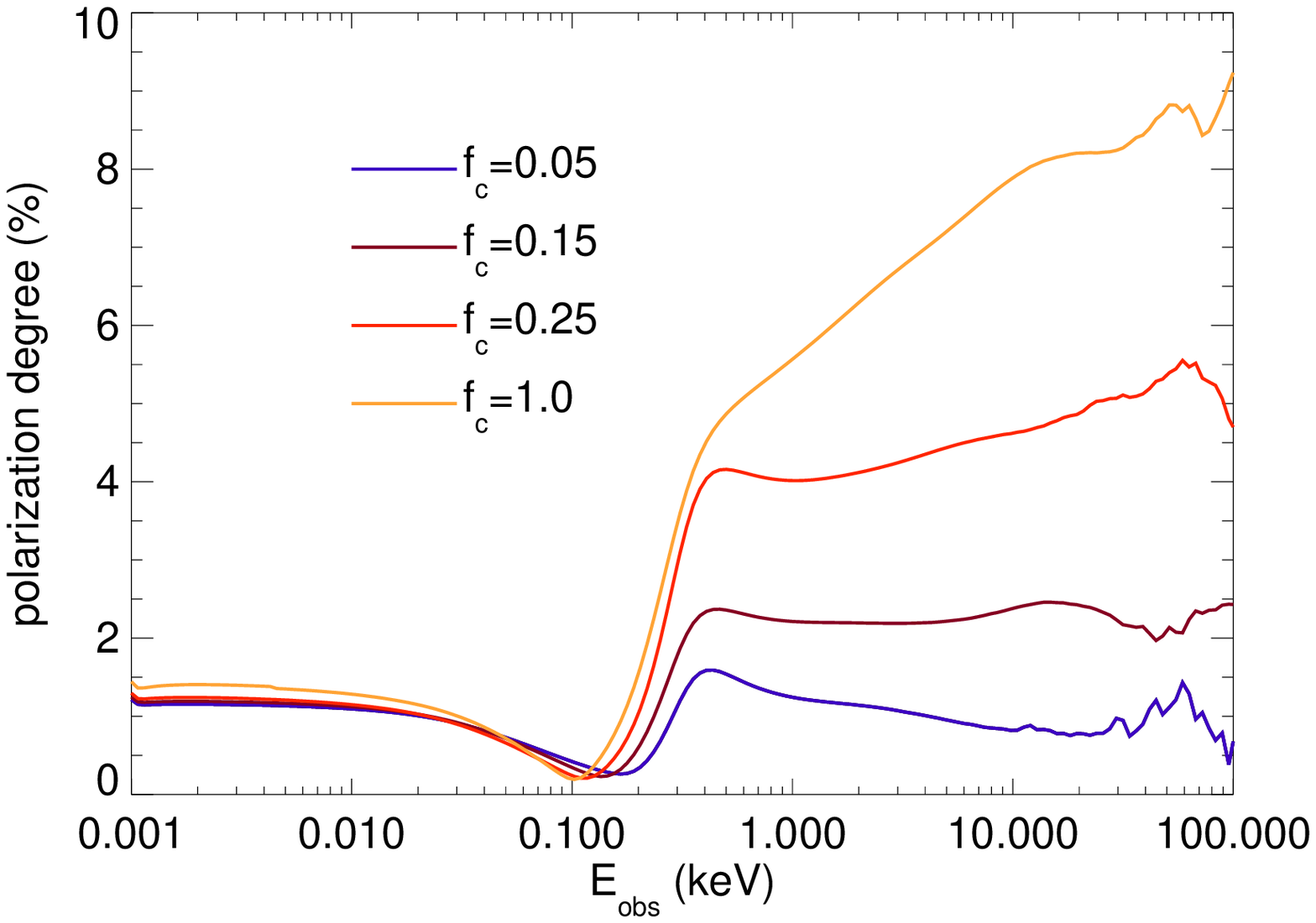}}
\scalebox{0.45}{\includegraphics{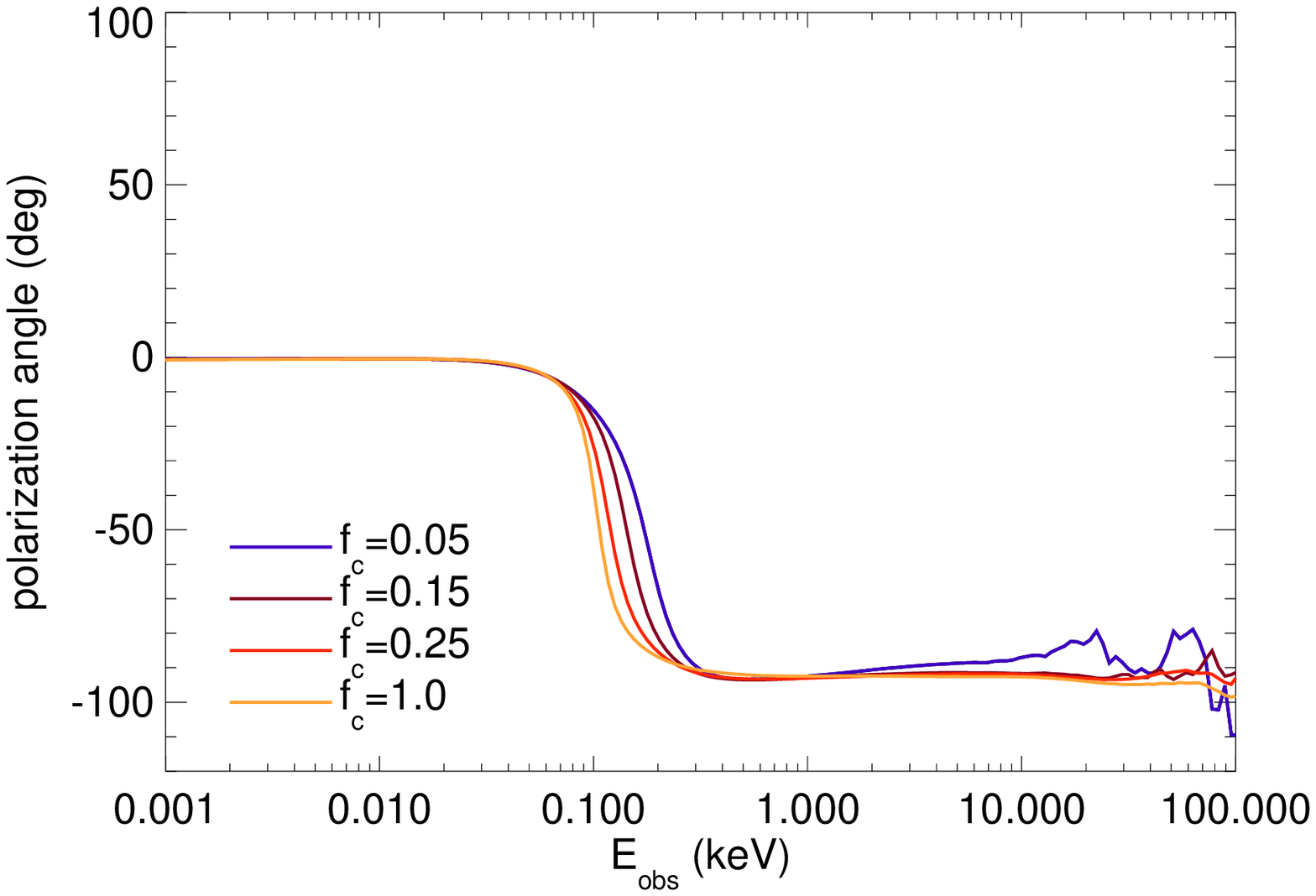}}
\end{center}
\end{figure}

As shown in \citet{schnittman:10}, polarization corollates closely
with covering fraction of the corona. In Figure \ref{agn} we show the
polarization from an AGN with central mass $M=10^7 M_\odot$, spin
parameter $a/M=0.9$, thermal luminosity $L_{\rm therm}=0.1 L_{\rm
  Edd}$, and observer inclination angle $i=45^\circ$. Four different
corona models are considered: a homogeneous wedge with $H/R=0.1$,
$\tau_0=1$, and $T_c=100$ keV, and three
clumpy models with the same scale height, mean optical depth, and temperature,
but with partial covering fractions of $f_c=0.25$, $0.15$, and
$0.05$. In the clumpy coronas, the density was chosen so that the
total flux in $2-10$ keV was approximately the same in all
cases. Again, the polarization transitions from horizontal to vertical
near the thermal peak, here around 100 eV. Clearly, the more uniform
coronas lead to higher polarization, 
again showing that polarization is fundamentally a measure of
symmetry. 

Like their stellar-mass counterparts, many AGN produce powerful
relativistic jets. In particular, the blazar class of AGN have their
jet axes pointing towards the observer, which leads to intense beaming
of radiation and rapid variability (see \citet{krawczynski:12c} for a
detailed discussion of blazar science with {\it GEMS}). Blazar spectra
are typically
characterized by two peaks: a synchrotron peak in the UV/optical/mid-IR, and
a secondary peak in the X-ray/gamma-ray band, likely caused by
the inverse-Compton scattering of either the synchrotron flux
\citep{poutanen:94,ghisellini:98} or possibly photons from the disk
\citep{mcnamara:09}. As mentioned above, polarization is most
sensitive to the scattering geometry, and thus is an ideal tool to
distinguish between internal and external inverse-Compton seeds
\citep{krawczynski:12a}. 
Since hot electrons generally cool faster via
synchrotron radiation than cool electrons, they will sample a smaller
and thus more coherent portion of the jet's magnetic field. It is
therefore quite likely that
X-rays created directly in the jet may be even more highly
polarized than the optical flux observed in many blazars
\citep{marscher:08}. However, in the synchrotron self-Compton picture,
it is the same electron population producing both the optical and
X-ray flux, so it is not obvious how the polarization will scale with
energy in those sources. 

The open questions regarding AGN corona and jets may be addressed with
X-ray polarization in
much the same way as the galactic BHs, with the noted
differences mentioned above, namely the lower-energy seeds require
more scatterings to reach the $2-10$ keV band, and the importance of
absorption of X-rays by partially ionized metals in the disk
atmosphere. On the practical level of connecting theory with
observation, and planning of new scientific missions, one must also
note that AGN are typically orders of magnitude less bright than
galactic sources, and are also more likely to be seen at lower
inclination angles where obscuration from surrounding dusk is
lowest, giving lower polarization amplitude for most models. On the
other hand, the persistence of even the most variable AGN makes them
more reliable targets than the transient galactic
sources. Furthermore, many of the bright nearby AGN also have an enormous
wealth of previous deep observations across the spectrum, including
polarization measurements in the radio, IR, and optical, all of which
bring valuable independent insight into the geometry of these
sources, and provide powerful constraints for many theoretical
models.

\section{GEMS TARGETS}
\label{source_simulation}

There are currently about 30 known galactic BHs
and BH candidates \citep{remillard:06}. The majority of these objects
spend most of their time in the quiescent state, but when they do go
into outburst, they are some of the brightest X-ray sources in the
sky, and thus ideal targets for photon-limited measurements like
polarization. Many of the simple models discussed above predict
polarization signatures that are strongly dependent on parameters like
the accretion disk inclination, mass accretion rate, BH mass, and the
distance to the source. All of these parameters could in principle be
determined with complementary observation in other wavelengths. Thus
it will be critical when trying to model real sources that we are able
to incorporate as much prior information as possible, in order to
maximize the relative value added by the polarization measurement.

Here we present a list of potential {\it GEMS} targets with a brief
summary of the observational properties of each source, and the
primary science questions that may be answered by each object. We also
include some suggestions for multi-wavelength observations, either
contemporaneous or in some cases, simultaneous. These targets are
summarized in Table \ref{target_list}. 

\subsection{Galactic BH binaries}
{\it Cyg X-1}: The first stellar-mass black hole discovered, and one
of the nearest and brightest in the galaxy. Recent observations have
provided exquisite measurements of its mass, distance, and inclination
\citep{reid:11, orosz:11}, which will make it even more valuable for
using polarization to measure spin. However, the relatively low
inclination suggests that we should expect low inherent polarization. 

{\it GX 339-4}: A relatively nearby source with numerous observations
of the iron line. The small mass leads to higher thermal temperatures,
which is useful for observing the thermal state with {\it GEMS'}
2-10 keV bandpass. The moderate yet uncertain inclination may make
it difficult to generate a high degree of polarization. 

{\it LMC X-3}: Despite its distance, and thus low flux, this is a
high-priority source due to its high duty cycle in the thermal state,
and relatively high inclination. With enough integration time, we
should be able to observe directly the effects of spin and strong
gravity on the polarization from a thin disk. 

{\it LMC X-1}: Similar to LMC X-3, but lower inclination, so lower
expected degree of polarization.

{\it GRS 1915+105}: This is an extremely bright, persistent source
that is highly variable and undergoes multiple state transitions,
predominantly in the hard and steep power low states, and exhibits a
rich selection of QPOs. It is a prime candidate for using
time-resolved polarization to study the accretion geometry of the
system. The large inclination makes high polarization likely. 

{\it 4U 1957+11}: A relatively faint source that may be in the
galactic halo. Mass and distance are not well-constrained, but thermal
spectrum is consistent with very rapidly spinning BH
\citep{nowak:11}. Inclination not well known, but likely high, so this
is a potentially promising source for testing thin disk models with
extreme spins.

\subsection{Active galactic nuclei}
{\it MCG 6-30-15}: The paradigmatic Seyfert-1 source for strong,
relativistically broadened iron line emitted from the inner disk. The
relatively high 2-10 keV flux makes it feasible for polarization
measurements, and the inclination is well-contrained by the
blue-shifted edge of the iron line. While this inclination is small
($\sim$30 degrees), the AGN corona models described in
\citet{schnittman:10} predict at least a few percent polarization in
this case. 

{\it NGC 1068}: A nearby Type-2 Seyfert galaxy with an edge-on disk, the X-ray flux could
potentially come from reflection of the central compact source off of
the surrounding hot wind, which would lead to a very high degree of
polarization \citep{antonucci:85}. The low flux will make it difficult to measure anything
but the highest polarization.

{\it Cen A}: Nearby bright FR I radio galaxy with large scale relativistic
jets. Galactic disk nearly edge-on, dominated by dust, gas, and star
formation. Good source for studying jets; central engine heavily
obscured below $\sim 5$ keV \citep{markowitz:07}. There is a strong
narrow Fe line and the continuum is modestly variable.

{\it NGC 4151}: A nearby Type-1 Seyfert with relatively large
absorption column \citep{weaver:94}. Recent observations suggest a
time lag between the continuum and a broad iron line, with the
continuum leading by ~2000 s \citep{zoghbi:12}. Combined with X-ray polarization, this
promises to be a powerful probe of the illumination geometry of the
inner disk. 

{\it NGC 5548}: A Sy 1.5 galaxy with estimated inclination $\sim$ 30
deg, but evidence for a disk truncated at large radius, as inferred
from the lack of a broad Fe line \citep{brenneman:12}. Could be an
excellent source for testing the spherical corona model for a
radiatively inefficient accretion flow.

\begin{table} 
\caption{\label{target_list} {\it GEMS} target list of galactic BHs
  and AGN. The fluxes are in the 2-10 keV band. The hard/steep power
  law states are designated ``H/SPL,'' and ``TD/VH'' represents the
  thermal-dominant/very high state. The duty cycles and fluxes come
  from the all-sky monitor on {\it RXTE}.}
\begin{tabular}{lcccccc}
\hline
Source      & Mass       & Distance & Inclination & State & Flux    & Duty Cycle \\
            & ($M_\odot$) & (kpc)   & (deg)       &       & (mCrab) &(\%) \\
\hline
GX 339-4    & 7$\pm$0.2  & 8$\pm$1  & 46$\pm$8    & H/SPL & 58      & 32\\
            &            &          &             & TD/VH & 290     & 25\\
GRS 1915+105& 14$\pm$4   & 11--12   & 66$\pm$2    & H/SPL & 750     & 95\\
            &            &          &             & TD/VH & 360     & 5\\
Cyg X-1     & 14.8$\pm$1 & 1.9$\pm$0.1 & 27$\pm$1 & H/SPL & 360     & 90\\
            &            &          &             & TD/VH & 880     & 10\\
LMC X-1     & 10.9$\pm$1.5 & 48$\pm$2 & 36$\pm$2  & H/SPL & 21      & 30\\
            &            &          &             & TD/VH & 21      & 68\\
LMC X-3     & 11.5$\pm$2 & 48$\pm$2 & 60$\pm$10   & H/SPL & 17      & 28\\
            &            &          &             & TD/VH & 25      & 65\\
4U 1957+11  & $>$3       & $>$10    & $<$75       & H/SPL & 32      & 28\\
            &            &          &             & TD/VH & 31      & 70\\
\hline
MCG 6-30-15 & 5$\times 10^6$ & 30,000 & 30        & SPL? & 2-3     & 100\\
NGC 1068    & 1.7$\times 10^7$ & 14,000 & $>$70   & ?   & $<$1     & 100\\
Cen A       & 2$\times 10^8$ & 3,500 & $\sim$60 & hard?     & 15   & 100\\
NGC 4151    & 4.5$\times 10^7$ & 14,000 & 45      & SPL?  & 13      & 100\\
NGC 5548    & 6.7$\times 10^7$ & 72,000 & 30      & hard  & 3       & 100\\

\end{tabular}
\end{table}

\section{EPILOGUE}
\label{epilogue}
{\it GEMS} was proposed in response to the NASA SMEX announcement of
opportunity  in December 2008, was selected for phase A development in
2009 and selected for phase B in 2010. A technically successful
Preliminary Design Review was held in Feb 2012. NASA Science Mission
Directorate (SMD) indicated their intention to non-confirm (or
cancel) in May 2012; the SMD decision was based on concerns that the
eventual cost would be too high.   

\newpage

\end{document}